\documentclass[prb, twocolumn]{revtex4-1}
\usepackage{amsmath,amssymb,bm}
\usepackage{graphicx}
\usepackage{epstopdf}
\usepackage{latexsym}
\usepackage{subfigure}
\usepackage{color}
\usepackage{natbib}
\usepackage{hyperref}
\usepackage{braket}
\hypersetup{
  colorlinks,
  citecolor=magenta,
  linkcolor=blue,
  urlcolor=blue}
      
\begin{document}

\title{Quadrupolar quantum criticality on a fractal}
\author{Jonathan D'Emidio}
\affiliation{Department of Physics \& Astronomy, University of Kentucky, Lexington, KY-40506-0055}
\author{Simon Lovell}
\affiliation{Department of Physics \& Astronomy, University of
  Kentucky, Lexington, KY-40506-0055}

\author{Ribhu K. Kaul}
\affiliation{Department of Physics \& Astronomy, University of Kentucky, Lexington, KY-40506-0055}
\begin{abstract}
We study the ground state  ordering of quadrupolar ordered $S=1$
magnets as a function of spin dilution probability $p$ on the
triangular lattice. In sharp contrast to the
ordering of $S=1/2$ dipolar N\'eel
magnets on percolating clusters, we find that the quadrupolar magnets are
quantum disordered at the percolation threshold, $p=p^*$. Further we
find that long-range quadrupolar order is present for all $p<p^*$ and
vanishes first exactly at $p^*$. Strong evidence
for scaling behavior close to $p^*$ points to an
unusual quantum criticality without fine tuning that arises from an interplay of quantum fluctuations and randomness.
\end{abstract}
\maketitle

\section{Introduction}

Quantum spin models with random vacancies provide a rich
playground for the study of the combined effects of strong interactions and
quenched disorder that are relevant to experimental measurements on a
number of doped magnetic
alloys~\cite{vojta2013:vietri}. Magnetic order in higher dimensions is generally stable to a small concentration of vacancies. On the other
 hand, if its moments are diluted beyond the percolation
 threshold, the system breaks up into zero-dimensional clusters and
hence magnetic order must be lost. What is the nature of the
 magnetic quantum phase transition as the dilution is varied? 

\begin{figure}[!t]
\centerline{\includegraphics[angle=0,width=0.8\columnwidth,trim=0 200 150 0,clip=true]{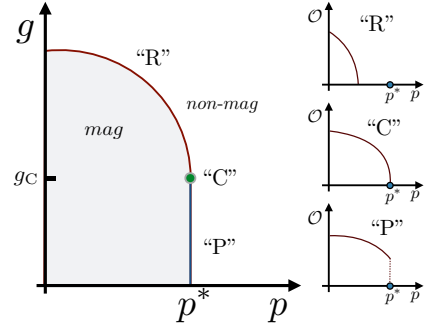}}
\caption{\label{fig:intro} Cartoon $T=0$ phase diagram for transverse
  field Ising model in the
  dilution probability $p$ and transverse field $g$ plane. For the
  largest cluster on a two-dimensional lattice occupied with
  probability $p$, the
  transition from magnetic to non-magnetic phases can be of three
  types as illustrated for the Ising order parameter $\cal{O}$
  vs. $p$: ``R''  the magnetic transition occurs before the percolation
  threshold and is hence expected to be identical to the random
  transverse filed Ising model. ``P'' the percolation quantum
  transition where the magnet remains ordered at the percolation
  threshold. ``C''  at which the order parameter
  vanishes continuously at $p^*$ achieved by fine tuning $g$ to a special value
  $g_{\rm C}$. For the $S=1$ quadrupolar magnets discussed here, we present
  detailed evidence that
the order vanishes like ``M'', but with no fine tuning,
leading to an unusual quantum criticality on a fractal cluster.}
\end{figure}

The answer to this question is most thoroughly understood for
the randomly diluted transverse field Ising
model. The expected ground state phase diagram~\cite{harris1974:ising}
as a function of the transverse field $g$ and
site dilution probability $p$, has three kinds of
quantum phase transitions, Fig.~\ref{fig:intro}. For $g>g_{\rm C}$ the
transition takes place at a value of $p$ smaller than the percolation threshold $p_c$, the critical
point is then described by the random transverse field Ising model
which possesses an infinite randomness fixed point~\cite{pich1998:ising,motrunich2000:ising}; the physics of percolation
however plays no role (``R'' in Fig.~\ref{fig:intro}). For
$g<g_{\rm C}$ on the other hand percolating clusters are magnetically ordered and the
singularities at the transition are determined by those of classical
percolation (``P'' in
Fig.~\ref{fig:intro})~\cite{senthil1996:perc}. 
Finally exactly at $g=g_{\rm C}$ a critical point is obtained: Crucial to our discussion, quantum criticality
at the percolation threshold  (``C'' in Fig.~\ref{fig:intro}) requires {\em fine tuning} of
the quantum fluctuations and hence ``C'' is not the generic transition for
Ising magnets on dilution.

Another family of quantum spin models where this question has been
addressed in detail are bipartite N\'eel ordered Heisenberg models. For random
depletion of the square lattice it was found that percolating clusters
have long range order~\cite{sandvik2002:s12perc}, the destruction is thus
akin to Fig.~\ref{fig:intro} ``P''. The surprising stability of
the N\'eel order has been traced back to the effect of uncompensated Berry phases
which result in ``orphan'' moments~\cite{wang2010:perc,changlani2013:perc}.  In the bilayer geometry,
with bond dilution that circumvents random Berry phases, a
phase diagram similar to Fig.~\ref{fig:intro} was found~\cite{vajk2002:bil,sandvik2002:bil}, with quantum
phase transitions corresponding to ``P''~\cite{vojta2005:perc} in some regions and
corresponding to ``R'' in others, in contrast to the Ising case the critical phenomena here
is controlled by a finite disorder fixed point with conventional scaling~\cite{sknepnek2004:heis}. The critical point ``C'' which
requires fine tuning has also been studied~\cite{sandvik2006:bilqc}. A
variety of phase diagrams and critical phenomena can be accessed by
dilution of N\'eel magnets in geometries different from the single and
bilayer~\cite{sandvik2006:bilqc,yu2005:s12dimperc}.

The two examples of dipolar ordered magnets (Ising and Heisenberg)
make it clear that the details such as symmetry of the order parameter
and Berry phases play a crucial role in
determining how quantum magnetism is destroyed by random dilution. In this
work we address the dilution transition in a different kind of
system: quadrupolar
ordered (also referred to as spin nematic) magnets in their most common $S=1$ realization~\cite{penc2011:quad}. The study
of the quadrupolar phase of $S=1$ magnets has become increasingly
popular motivated by their possible sighting in some triangular lattice
Ni based magnets (see e.g.~\cite{nakatsuji2005:nigas,tsunetsugu2006:sn,bhattacharjee2006:sn,lauchli2006:quad,stoudenmire2009:quad,smerald2013:sn}). Experimental studies of Zn replacement of Ni
in NiGa$_2$S$_4$ provide a direct experimental motivation for the site
diluted $S=1$ magnets we study here~\cite{yusuke2008:znnigas}. As we
shall describe below, in contrast to what has been
observed for N\'eel order in $S=1/2$ magnets and transverse field Ising models, we
find here that quadrupolar order vanishes exactly at the percolation threshold {\em without any
fine tuning}, resulting in new quantum critical behavior.

\begin{figure}[!t]
\centerline{\includegraphics[width=\columnwidth]{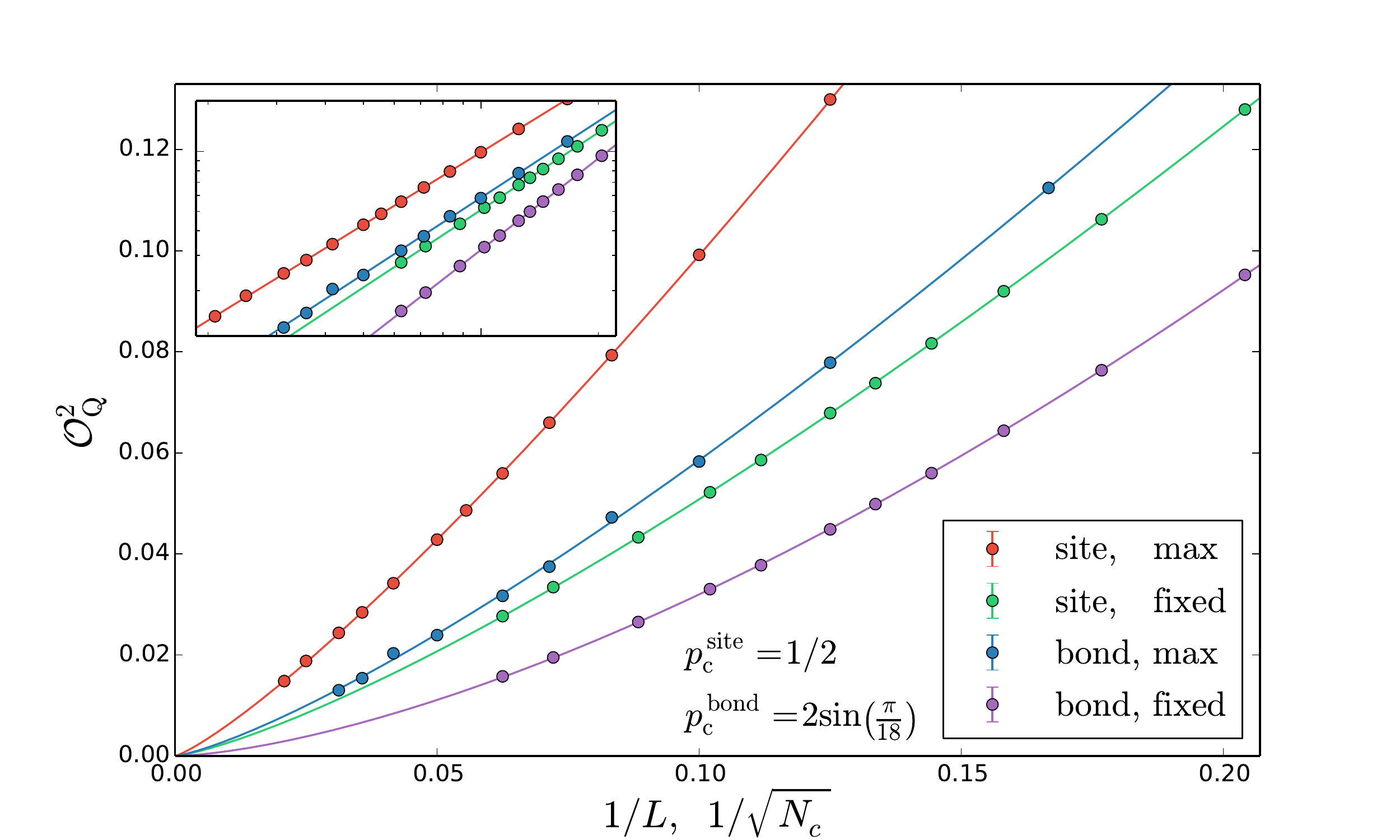}}
\caption{\label{fig:orderpc} Thermodynamic extrapolation of the $T=0$
  quadrupolar order parameter ${\cal O}_Q^2$ at the percolation threshold. We present
  results of  simulations of the model Eq.~(\ref{eq:s1btri}) using both site diluted and bond diluted triangular lattices at
their respective $p^*$. For each case we have carried
out the averages for finite-size scaling in two ways: using an
ensemble with a {\em fixed}
cluster of size $N_c$ or an ensemble of the
largest {\em max} cluster on an $L\times L$ lattice. In all four cases, we find
${\cal O}_Q^2$ vanishes in the thermodynamic limit at $p^*$. The inset
shows the same data on a log-log scale. }
\end{figure}

\section{Model}

Just as the bilinear Heisenberg model is the archetype for
realizing N\'eel order, the biquadratic Heisenberg interaction is the
archetypical model system that realizes quadrupolar order in $S=1$
magnets. We shall consider the following Hamiltonian,
\begin{equation}
\label{eq:s1btri}
 H = - \sum_{\langle ij\rangle} J_{ij} \left (\vec S_i \cdot \vec S_j\right )^2,
\end{equation}
 where $\langle ij \rangle$ denote nearest neighbors in the triangular
 lattice, and $\vec S_i$ are the $S=1$ Pauli matrices on the $i^{\rm
   th}$ site. The model possesses an explicit physical SO(3) internal spin
 symmetry.  We use standard uncorrelated random site and bond
 dilution: For site dilution $J_{ij}=|J| r_i r_j$ where $r_i=0$ with
 probability $p$ and $r_i=1$ with probability $1-p$. For bond dilution
 $J_{ij}=|J| \chi_{ij}$ where $\chi_{ij}=0$ with probability $p$ and
 $\chi_{ij}=1$ with probability $1-p$. While site dilution is closer
 to experimental realizations such as Zn doped NiGa$_2$S$_4$,
 bond dilution provides an alternate access to the percolation
 threshold. 

Our numerical results were obtained
 using stochastic series expansion~\cite{sandvik2010:vietri} with an efficient sign-problem free algorithm
 that has been described and tested
 previously~\cite{kaul2012:quad}. We calculate averages on finite size
 clusters at finite-temperature, $T$. 
To study the quadrupolar order in the magnet we define the quadrupolar
order parameter through the equal time two point
correlation function of the quadrupolar order parameter.

\begin{equation}
\label{eq:op}
{\cal O}_Q^2= \langle \frac{1}{N_c^2} \sum_{a,i,j} \langle \hat
Q^{aa}_i \hat Q^{aa}_j\rangle \rangle_p,
\end{equation}
where the quadrupolar order parameter is defined in terms of the local $S=1$
Pauli matrices as
$\hat Q^{ab}= \frac{\hat S^a \hat S^b+ \hat S^b\hat S^a}{2}-
\frac{2}{3}\delta^{ab}$, $N_c$ is the size of the cluster, $\langle \dots \rangle$ is the quantum
mechanical average over the thermal density matrix and $\langle \dots
\rangle_p$ is a disorder average over realization of the ensemble of clusters.

\section{ Numerical Results}

It is now well established that the model
Eq.~(\ref{eq:s1btri}) without any depletion ($p=0$) has quadrupolar
order with ferromagnetically aligned directors~\cite{lauchli2006:quad,
kaul2012:quad,voll2015:quad} (i.e. ${\cal O}_Q^2$ in Eq.~\ref{eq:op}
is finite in the ground state in the thermodynamic limit). We begin our numerical study by asking the
following question: As $p$ is increased does the quadrupolar order parameter
eventually vanish like ``R'', ``C'' or ``P'' (referring to Fig.~\ref{fig:intro}).  As we have discussed in
the introduction,
generic expectations would be
either ``R'' or ``P''.

\begin{figure}[!t]
\centerline{\includegraphics[width=\columnwidth]{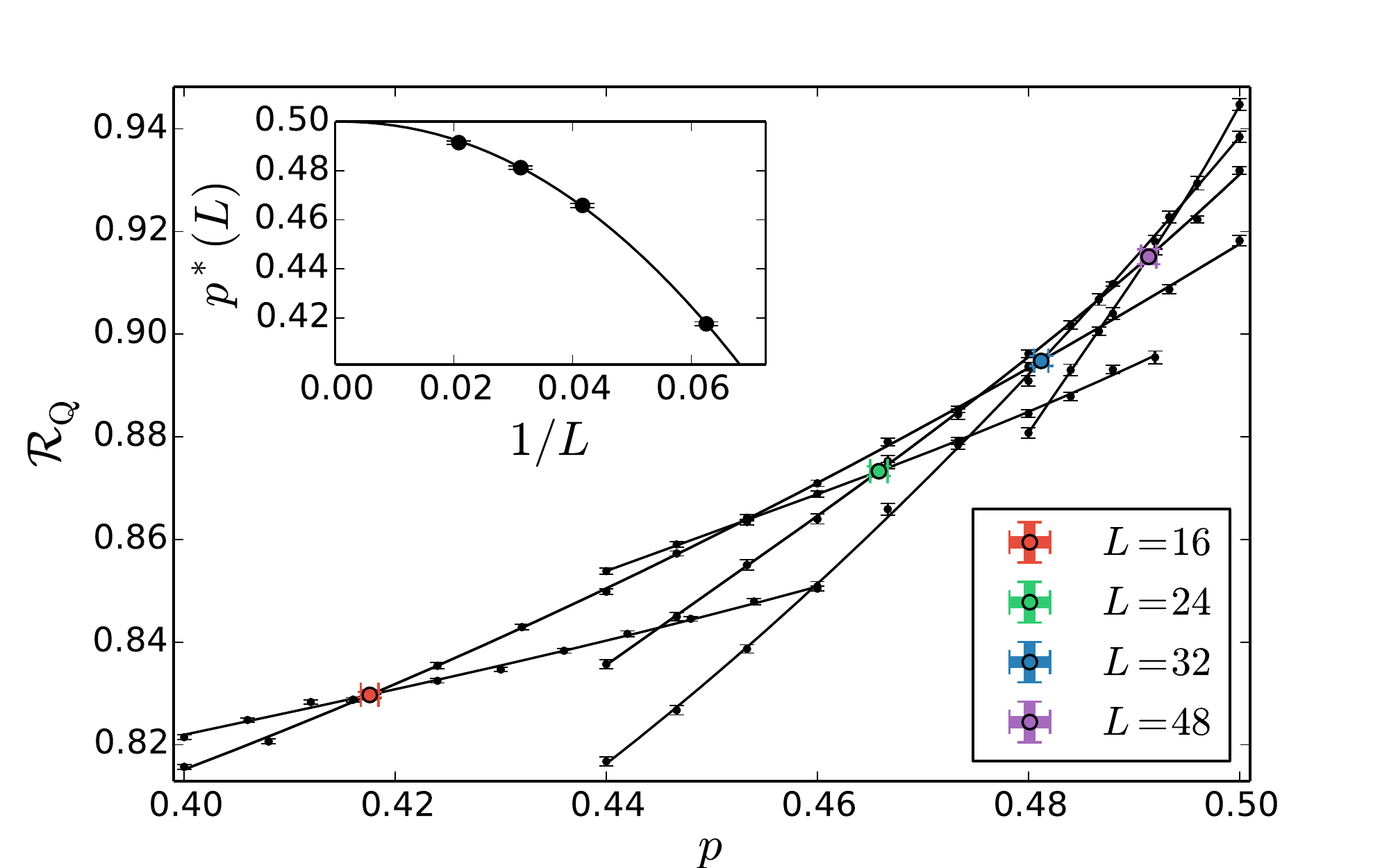}}
\caption{\label{fig:binder} Crossing of the Binder ratio ${\cal R}_Q$
  for the site diluted triangular lattice, demonstrating that
  the order parameter vanishes first at $p=p^*$ with high precision. The
  main panel shows a zoom in of ${\cal R}_Q$ as a function of the site occupation
  probability $p$ close to the percolation threshold. The crossings points
  between $(L/2,L)$ pairs are marked with errors.  Inset
  shows the $p(L)$ value at the crossings of $(L/2,L)$ pairs. The crossing
  data extrapolates accurately to the
  percolation threshold of the site diluted triangular lattice, $p^*=0.5$ with a power law fit. This numerical evidence establishes that the
  quadrupolar order parameter is finite for all $p<p^*$ and vanishes
  first exactly at the percolation threshold as illustrated in case
  ``M'' in Fig.~\ref{fig:intro}. }
\end{figure}

In Fig.~\ref{fig:orderpc} we show extrapolations of the quadrupolar
order parameter to
the thermodynamic limit at the percolation threshold. We define two different ensembles for
finite-size scaling which allow us to approach the thermodynamic limit in two
different ways, with fixed cluster size $N_c$ (on an infinite
underlying lattice) and the largest cluster on a finite $L\times L$
lattices, following~\cite{sandvik2002:s12perc}. We have also studied
both bond and site dilution at the percolation threshold. Consistently across
all finite-size scaling schemes we find that the order parameter vanishes
at the percolation threshold. For a similar scaling analysis for the
N\'eel order, see Fig. 10 in Ref.~\cite{sandvik2002:s12perc}
which clearly shows the ordering of $S=1/2$ Heisenberg model on the
percolating cluster in two dimensions. In contrast here we find the
quadrupolar order vanishes in the thermodynamic limit. This data clearly eliminates the
``P'' possibility since this requires the percolating clusters to be
magnetically ordered.   We have taken great care to make sure our
data is equilibrated and in the limit of $T=0$. The ground state limit
requires extraordinarily low temperatures becasue of the weak links
that connect percolating clusters. Details are provided
in the Appendix.

\begin{figure}[!t]
\centerline{\includegraphics[width=\columnwidth]{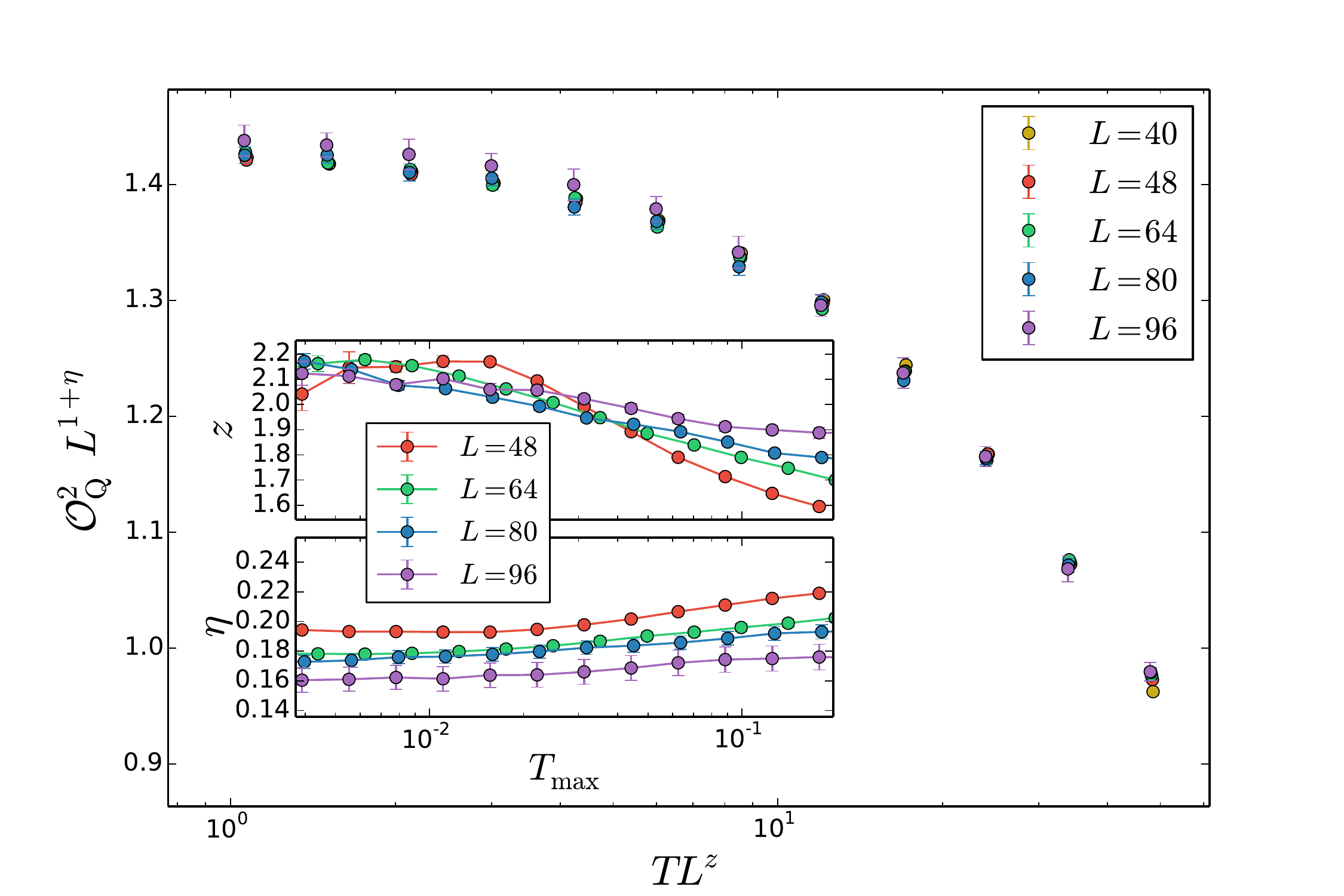}}
\caption{\label{fig:dynscale} Finite temperature scaling collapse of the quadrupolar order
  parameter at $p=p^*$. Here we are testing the quantum critical
  scaling form ${\cal O}_Q^2 = \frac{1}{L^{1+\eta}}{\cal F} ( L ^z
  T)$. The main panel shows the collapse of the order parameter at
  $p=p^*$, using the values $z=1.987$ and $\eta=0.178$ for the exponents. The inset illustrates the finite
  size corrections and convergence of the critical exponents: the
  values of $z$ and $\eta$ obtained from pair-wise collapse for $(L,L/2)$ are graphed as a function of
  the maximum temperature data used in the collapse. The drifts on the
  largest system sizes at the smallest temperatures are relatively
  small allowing us to make reliable estimates for the exponents and
  their error windows: $z=2.1(2)$ and $\eta=0.17(2)$.}
\end{figure}

Now that we have shown that the percolating cluster are magnetically
disordered, the generic expectation is that the quadrupolar order
vanishes before the percolation threshold is reached at some $p<p^*$
as illustrated for ``R''  in Fig.~\ref{fig:intro}. 
To address this question quantitively we study a quadrupolar
``Binder'' ratio ${\cal R}_Q\equiv\frac{\langle {\cal O}_Q^4
  \rangle}{\langle {\cal O}_Q^2 \rangle^2}$. ${\cal R}_Q $ is expected
to be monotonically {\em decreasing}
with $L$ in a quadrupolar ordered phase. When graphed as a function of the tuning
parameter $p$, ${\cal R}_Q$ data for different $L$ are expected to cross at the phase
transition to a non-magnetic phase. Our zero temperature data for the
case of site dilution in Fig.~\ref{fig:binder} clearly shows that the ${\cal
  R}_Q$ crosses and as the thermodynamic limit is reached  the crossing point approaches the percolation
threshold $p^*=0.5$ with high accuracy. This establishes that the
quadrupolar order vanishes continuously as the percolation threshold
is approached. This allows us to eliminate the
possibility ``R'', in which the order parameter vanishes before the
percolation threshold is reached. We hence conclude that the order
parameter vanishes first precisely at the percolation threshold $p^*$, as
illustrated in the cartoon ``C'' shown in Fig.~\ref{fig:intro}.

The vanishing of quadrupolar order right at the
percolation threshold raises the interesting possibility that at $p=p^*$ the
system is quantum critical and the quadrupolar correlation possess
scale invariance.

Random systems are well known to display a range of
novel scaling behavior. We present evidence however that 
our model possesses conventional power law scaling. In order to test the scaling hypothesis, we first study the dynamic scaling in
imaginary time of the order parameter at $p=p^*$. In
Fig.~\ref{fig:dynscale} we show scaling collapse of the order
parameter data at the percolation threshold $p=p^*$ for the maximum sized cluster
on an $L\times L$ lattice as a function of
the putative scaling variable $L^zT$. For a quantum critical system,
we expect the scaling form ${\cal O}_Q^2 = \frac{1}{L^{1+\eta}}{\cal F} ( L ^z
  T)$. Note the absence of a tuning parameter (we only require the
  system to be at the percolation threshold) that would normally be
  expected for instance
  from previously studied phase diagrams of quantum rotor models in
  Fig.~\ref{fig:intro} (the tuning parameter there is the choice of
  $g=g_{\rm C}$). We find an excellent collapse with small finite size
  corrections to the exponents.

\begin{figure}[!t]
\centerline{\includegraphics[width=\columnwidth]{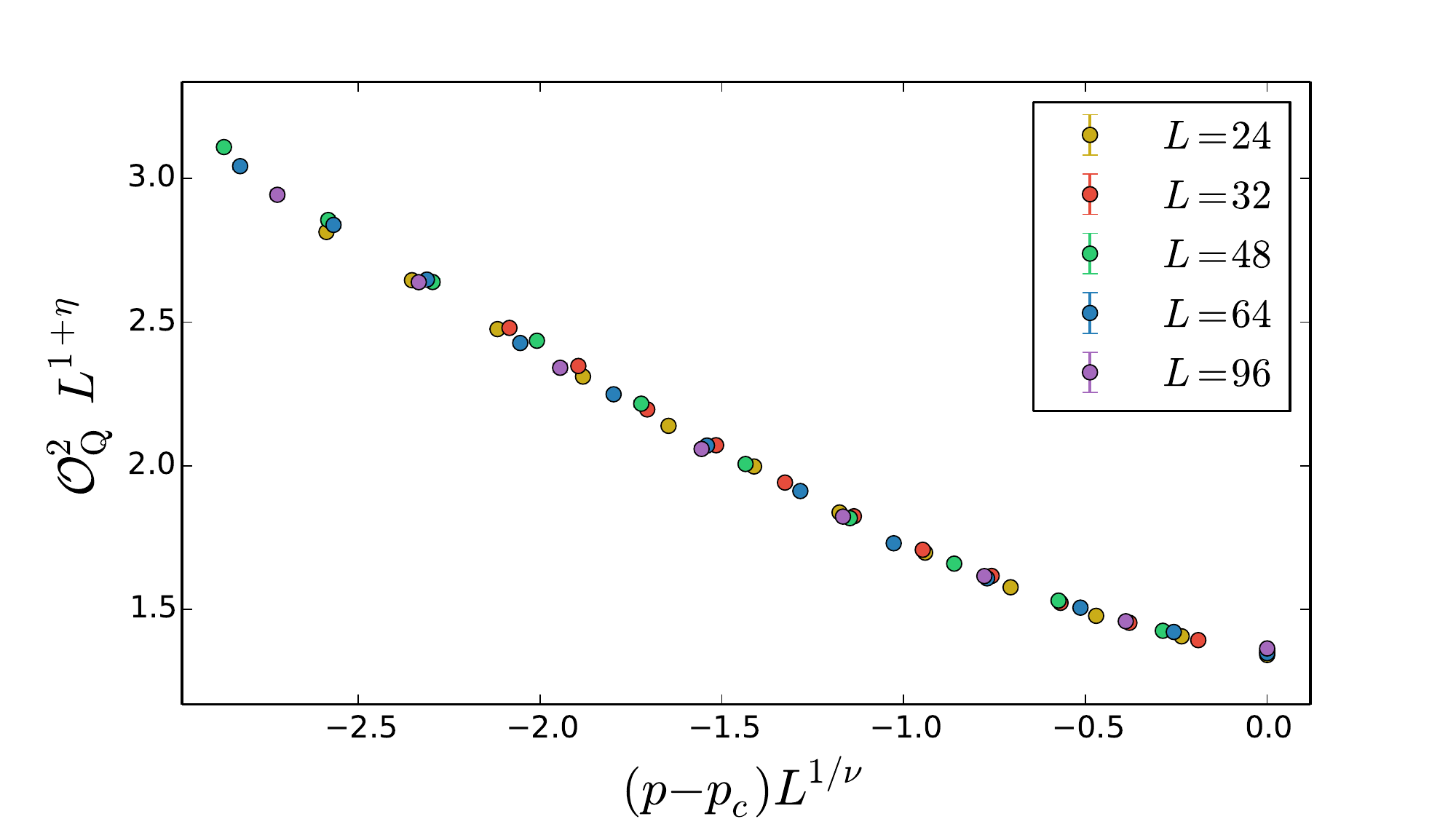}}
\caption{\label{fig:nu} Scaling collapse of the order
  parameter as a function of the deviation from the percolation
  threshold $p-p^*$ . A good collapse is found without any corrections
  to scaling with $\eta=0.196$ and $\nu=0.977$. We have collected this data assuming $z=2$ and hence
  fixed $T=1/L^2$. The critical exponents 
  were used for this collapse. From pair-wise collapse similar to
  Fig.~\ref{fig:dynscale} we find $\nu=1.01(5)$ and
  $\eta=0.18(1)$. The value obtained for $\eta$ is consistent with the
collapse in Fig.~\ref{fig:dynscale}.} 
\end{figure}

An alternate test of scaling can be made by varying $p$ away from the
percolation threshold. In Fig.~\ref{fig:nu} we study the scaling of
the order parameter with the deviation from the percolation theshold
$p-p^*$. We find excellent scaling behavior assuming the simple
scaling form ${\cal O}_Q^2 = \frac{1}{L^{1+\eta}}{\cal G} (
(p-p^*)L^{1/\nu})$ . The value of $\eta$ so obtained is in excellent
agreement with the estimate obtained from the previous scaling analysis 
at the percolation threshold Fig.~\ref{fig:dynscale}.

\section{Discussion}

We have presented extensive evidence that in two dimensions
quadrupolar order vanishes continuously as a function of dilution
right at the percolation threshold. At the percolation threshold the
systems shows conventional scaling behavior symptomatic of a finite
disorder fixed point. Our study is an unusual example of quantum
criticality at the percolation threshold without any fine tuning. We
contrast this with the generic phase diagram Fig.~\ref{fig:intro}
where at the percolation threshold, tuning of quantum fluctuations to
a special value if required to observe quantum criticality.

An interesting open question is whether the critical exponents we have
found are universal or can vary continuously as non-universal
properties of the magnet are varied at the percolation threshold. Such
varying exponents were reported in a study of specific diluted
dimerized $S=1/2$ magnet~\cite{sandvik2006:bilqc}. Interestingly, the dynamic critical exponent $z$ we find from our scaling analysis is
numerically somewhat close to
$d_f=91/48\approx 1.8958\dots$ (the fractal dimension of a percolating
cluster). A number of previous works have found or predicted such scaling at the
percolation threshold and our finding could be consistent with such
behavior~\cite{yu2005:s12dimperc,sandvik2002:bil,vojta2005:perc},
though numerically $z=2$ would also be consistent with our value
within errors. 
A complete theory of the unusual behavior and scaling we have found here is
an interesting direction for future work.

We acknowledge financial support fron NSF DMR-1611161. The numerical
simulations reported here were carried out on the DLX cluster at
University of Kentucky and by resources allocated by XSEDE.

\bibliography{perc}

\appendix

\section{Equilibration of disordered clusters}
\label{app:eql}

In order to ensure the proper equilibration of our disordered clusters, we have adopted the same equilibration and measurement protocol as in Ref.~\onlinecite{sandvik2002:s12perc}.  For each disorder realization we begin by performing $N_e$ equilibration sweeps at some large initial temperature ($\beta \approx 1$, setting $|J|=1$), followed by $N_m$ more measurements sweeps, then again $N_e$ equilibration sweeps, and finally $N_m$ measurement sweeps.  Once this process is complete we perform $\beta$ doubling on our configurations (see appendix \ref{app:zeroT}) and start the process again.  Thus for each disorder realization we have two separate measurement cycles at many different values of $\beta$.

\begin{figure}[!t]
\centerline{\includegraphics[angle=0,width=1.0\columnwidth]{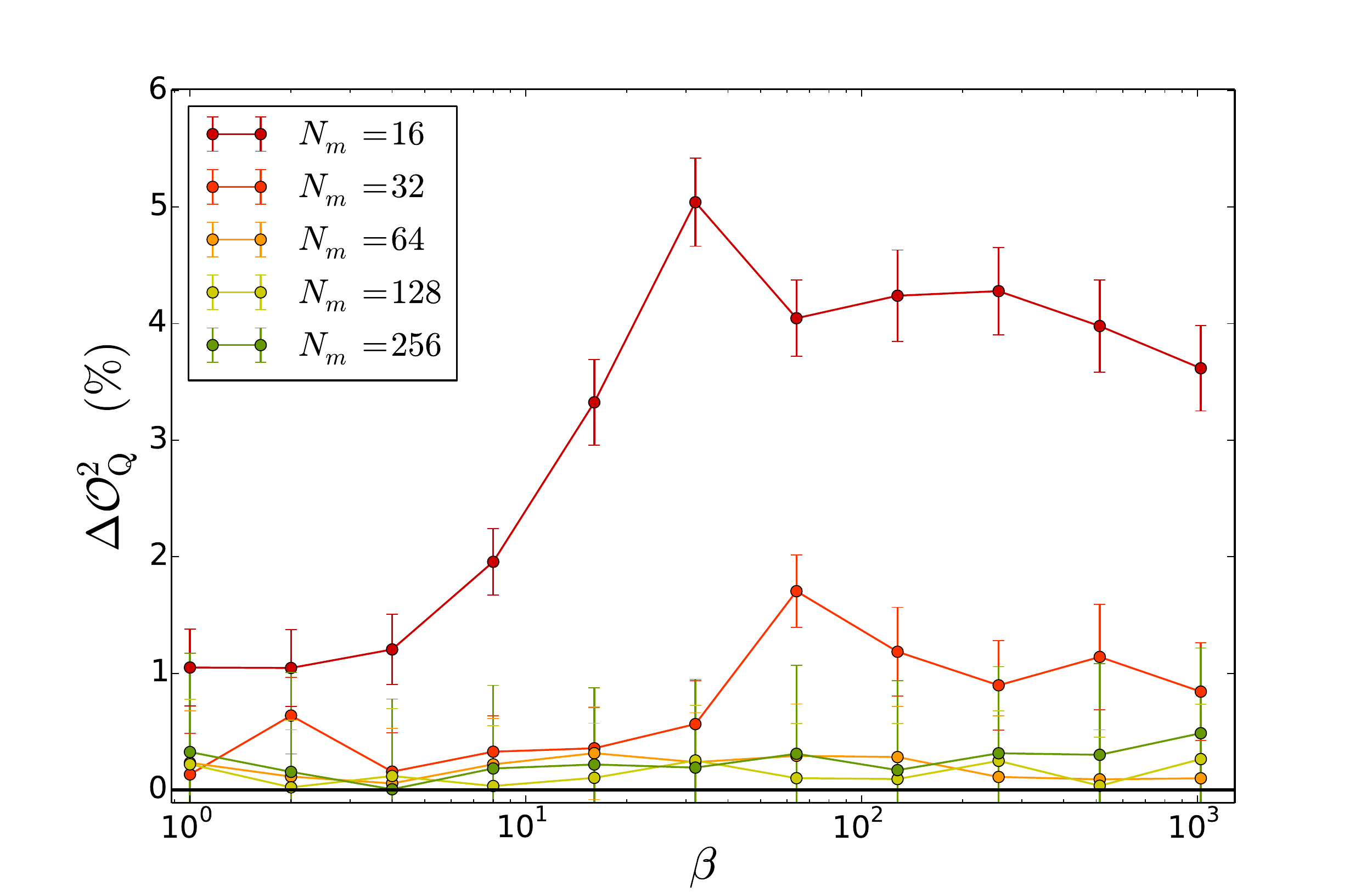}}
\caption{Here we show the percent difference of the quadrupolar order parameter between the first and second measurement segments ($\Delta \mathcal{O}^2_{\mathrm{Q}}$) as a function of $\beta$ for several different values of $N_m=2 N_e$.  The difference between the two segments becomes statistically insignificant over the whole temperature range near $N_m=64$, indicating that we have sufficiently equilibrated our configurations beyond this number of sweeps.  We therefore cautiously set $N_m=200$ throughout the course of our numerical studies.}
\label{fig:swpCon}
\end{figure}

Separating out two distinct measurement segments allows us to check the equilibration of our disordered clusters.  In order to achieve the best QMC averages with minimal computational time, we have set $N_m=2 N_e$.  We can then study the percent difference of our two measurement segments (each averaged over disorder realizations) as a function of $N_m$ to determine its optimal value, which is given in Fig. \ref{fig:swpCon}.  We observe that at the percolation threshold (where the clusters are most fragmented), the difference between our first and second measurement cycles becomes statistically insignificant when $N_m \approx 64$.  We therefore cautiously set $N_m=200$ throughout the course of our numerical studies.

\section{Zero temperature convergence}
\label{app:zeroT}

The most computationally expensive component of our numerical studies
is to converge our configurations to the ground state on disordered
clusters.  In order to achieve this, extremely large values of the
inverse temperature are needed (relative to that of a clean lattice).
Furthermore, the free energy landscape of configurations is rugged,
and thus quenching a randomly initialized starting configuration to
low temperature abruptly may leave it stuck in a local minimum.  In
order to circumvent this issue, and to efficiently reach the low
temperatures required with minimal equilibration, we implement the
$\beta$ doubling procedure~\cite{sandvik2002:s12perc}.  The procedure works as follows: given a disorder realization, we begin the equilibration and measurement cycles at some initial high temperature ($\beta_0$ on the order of unity).  Configurations at high temperatures are relatively easy to equilibrate, since there are few operators acting in the operator string.  After the equilibration and measurement cycles, the value of $\beta$ and the QMC configuration are both doubled.  In terms of the configuration, this corresponds to repeating the operator sequence twice.  This procedure gives another valid partition function configuration that is close to being equilibrated at an inverse temperature $2\beta_0$.  From here the equilibration and measurement sequence is again carried out.  The process is continued until the final target value $\beta_{\mathrm{max}}$ is reached.  

\begin{figure}[h]
\centerline{\includegraphics[angle=0,width=1.0\columnwidth]{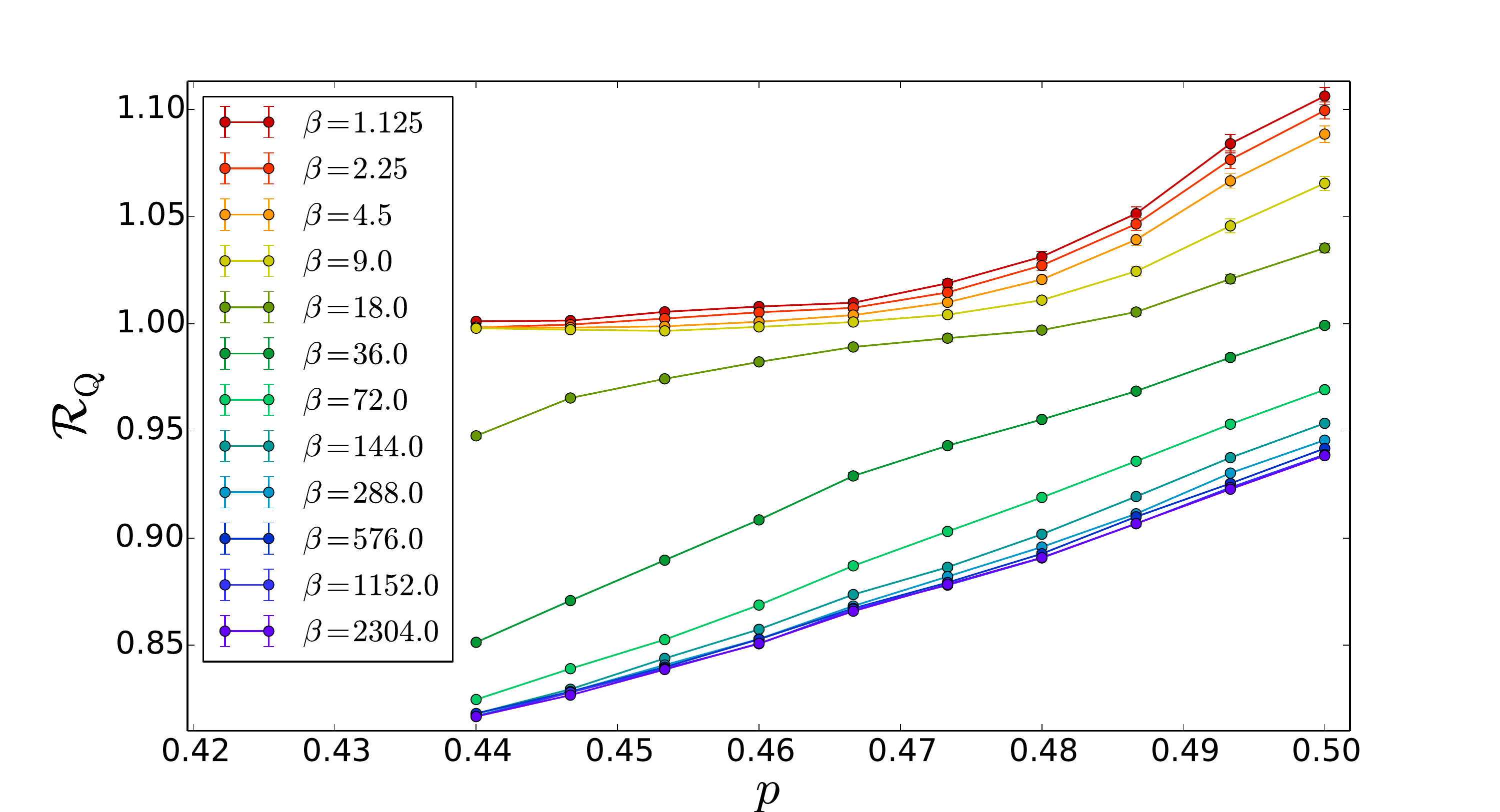}}
\caption{Here we show the zero temperature convergence of the quadrupolar binder ratio near the site diluted percolation threshold for max clusters on an $L=32$ lattice.  Each one of the $\beta$ values was obtained by the $\beta$ doubling procedure explained in appendix \ref{app:zeroT}.  We note that very large values of $\beta$ are required to converge to the ground state, especially near the percolation threshold ($p_c=0.5$) where we observe quantum critical behavior.}
\label{fig:zeroTbinder}
\end{figure}

In order to illustrate the convergence of our zero temperature data, in Fig. \ref{fig:zeroTbinder} we show the binder ratio for max clusters on an $L=32$ lattice near the site percolation threshold for all of our $\beta$ doubled values.  We see that very large values of $\beta$ are required to converge to the ground state.  This is a reflection of the fact that $z$ is close to $2$ at this disordered quantum critical point, meaning that doubling the size of the lattice would require quadrupling $\beta$ to remain near the ground state.

\end{document}